\documentclass[final,11pt]{article}
\usepackage{multirow}
\usepackage[nottoc]{tocbibind}
\usepackage{geometry}
\usepackage{subcaption}
\usepackage[affil-it, auth-sc]{authblk}
\usepackage[english]{babel}
\usepackage[utf8]{inputenc}
\usepackage{graphicx}
\usepackage{amssymb}
\usepackage{amsthm}
\usepackage{amsmath}
\usepackage{amsfonts}
\usepackage{verbatim}
\usepackage{lineno}
\usepackage{float}
\usepackage{todonotes}
\presetkeys{todonotes}{inline, color=blue!30, size=\small}{}
\usepackage{color, soul}
\definecolor{darkred}{rgb}{0.9, 0.0, 0.0}
\definecolor{darkgreen}{rgb}{0.0, 0.5, 0.0}
\usepackage[colorlinks,bookmarks,linkcolor=darkred, citecolor=darkgreen]{hyperref}
\usepackage{eso-pic}
\usepackage[sort&compress,numbers]{natbib}
\usepackage{doi}
\usepackage{paralist,epsfig}
\usepackage{relsize}
\usepackage{nicefrac,esint}
\usepackage{cleveref}
\emergencystretch=\maxdimen
\begin{document}

\AddToShipoutPictureFG*{
    \AtPageUpperLeft{\put(-60,-60){\makebox[\paperwidth][r]{LA-UR-21-30814}}}
}

\title{Broadening of particle distributions in electron- and (anti)neutrino-nucleus scattering from QED interactions}

\author{Oleksandr Tomalak\thanks{tomalak@lanl.gov}}
\author{Ivan Vitev\thanks{ivitev@lanl.gov}}
\affil{Theoretical Division, Los Alamos National Laboratory, Los Alamos, NM 87545, USA}

\date{\today}

\maketitle
Proper interpretation of past, current, and future data on lepton-nucleus reactions requires a clear separation between quantum electrodynamics (QED) and strong interaction effects inside the nucleus. First studies of QED in-medium lepton dynamics have set a theoretical framework to derive  electron-nucleus and (anti)neutrino-nucleus cross-section corrections. We employ this approach to quantitatively compute the effects of Glauber photon-mediated multiple re-scattering within the nuclear medium. We find that the relativistic charged lepton acquires momentum of order $10~\mathrm{MeV}$ transverse to its direction of propagation inside the nucleus. This broadening sizably deflects expected electron tracks and suppresses scattering cross sections. Precise extraction of the nucleon and nuclear structure by electron and muon probes should, thus, take the QED nuclear medium angular redistribution of particles into account. Our results further show that the associated effects in (anti)neutrino-nucleus scattering with measured final-lepton energy are significant only at the kinematical endpoints.

\newpage
\tableofcontents

\section{Introduction}

Lepton scattering off protons and nuclei is a traditional and time-honored way of studying the structure of hadrons and nuclear bound states. As an illustrative example, the nucleon form factors and nuclear spectral functions are extracted via electron scattering experiments with a percent-level or better precision~\cite{Dombey:1969wk,Akhiezer:1974em,A1:2010nsl,A1:2013fsc,Xiong:2019umf,Punjabi:2015bba,Bosted:1989hy,Perdrisat:2006hj,JeffersonLabHallA:2022cit,JeffersonLabHallA:2022ljj}. With significant progress in computation, the same nucleon form factors are also evaluated from the QCD Lagrangian on a space-time lattice~\cite{Gockeler:2003ay,Alexandrou:2018sjm,Yamazaki:2009zq,Chambers:2017tuf,Capitani:2015sba,Bruno:2014jqa,Bratt:2010jn,Kronfeld:2019nfb,Ishikawa:2018rew,Hasan:2019noy,RQCD:2019jai,Jang:2019vkm,Bali:2019yiy,Alexandrou:2020okk,Park:2021ypf,Djukanovic:2022wru,Jang:2023zts}. Besides providing fundamental constant values for nuclear physics, precise determination of vector and axial-vector structure of nucleons and nuclei is important for electroweak physics studies in the neutrino and charged lepton sectors of the Standard Model. New measurements and calculations are underway  to significantly reduce the dominant cross-section uncertainties in the extraction of neutrino oscillation parameters, in the determination of the neutrino mass hierarchy, and in searches for the CP-violation in the lepton sector with modern and near-future accelerator-based neutrino experiments~\cite{Nunokawa:2007qh,NOvA:2007rmc,T2K:2011qtm,KamLAND:2013rgu,MicroBooNE:2015bmn,JUNO:2015zny,Hyper-KamiokandeProto-:2015xww,McConkey:2017dsv,RENO:2018dro,DayaBay:2018yms,Machado:2019oxb,MicroBooNE:2019nio,DoubleChooz:2019qbj,Farnese:2019xgw,T2K:2019bcf,NOvA:2019cyt,DUNE:2020ypp,JUNO:2021vlw}.

To precisely and unambiguously determine the process-independent nucleon and nuclear structure from the experimental data, we have to unfold the radiative corrections according to the experimental setup. These radiative corrections and detector responses are very different in (anti)neutrino and electron scattering measurements at GeV energies and below. While significant amount of work has been dedicated to electron-proton and electron-nucleus scattering~\cite{Yennie:1961ad,Mo:1968cg,Maximon:2000hm,Vanderhaeghen:2000ws,Gramolin:2014pva,Afanasev:2023gev}, consistent treatment of charged-current (anti)neutrino-nucleus processes is a relatively new field~\cite{DeRujula:1979grv,Day:2012gb,Tomalak:2021hec,Tomalak:2022xup}. At GeV energies, radiative corrections in electron scattering experiments are typically at the $10$-$20\%$ level and slightly smaller in neutrino scattering. Besides radiative corrections, the exchange of photons between charged particles and the nuclear medium in (anti)neutrino-, electron-, and muon-induced processes inside a large nucleus modify the cross sections of the elastic scattering off nucleons.\footnote{By elastic scattering on nucleons, we denote the process with initial and final states consisting of the lepton and the nucleon and allow for change of the kinematics.} Such QED medium effects were found to be at the permille level in (anti)neutrino-nucleus scattering and up to percent level in electron-nucleus scattering in forward kinematics~\cite{Tomalak:2022kjd}.

In this work, we continue the evaluation of QED medium effects and study phenomenological consequences of the multiple re-scattering of charged leptons in a large nucleus with an emphasis on close connection to the ways in which electron and neutrino scattering experiments are performed. In the case of quantum chromodynamics (QCD) the re-scattering of quarks and gluons in nuclear matter~\cite{Accardi:2002ik,Accardi:2012qut,Gyulassy:2002yv} has attracted attention as a possible explanation of the Cronin effect~\cite{Antreasyan:1978cw}. It was realized early on that such soft interactions in matter will also lead to angular de-correlation in the distribution of final state particles in reactions with nuclei~\cite{Qiu:2003pm,Kang:2011bp}. This research continues today for various types of nuclear matter~\cite{Chen:2016vem,Barata:2020rdn}. In contrast, the QED effects mediated by Glauber photons remain practically unexplored. We begin to address this knowledge gap by quantifying  the corresponding changes on the experimentally accessed electron- and (anti)neutrino-nucleus cross sections and study the deflection of the charged lepton trajectories by multiple soft interactions within the nucleus before and after the hard scattering on nucleons. We find QED nuclear medium-induced redistribution of particles in the transverse direction that is sizable and important for the interpretation of the measurements of nucleon and nuclear structure. The angular deflection of electron tracks inside the nucleus and induced cross-section corrections should be verified experimentally in electron-nucleus scattering and taken into account in modern and future extractions of nuclear and QCD structure quantities.

The rest of our paper is organized as follows. In Section~\ref{sec:formalism}, we discuss multiple re-scattering of charged leptons via Glauber photon exchanges and present the results for the transverse momentum distribution of outgoing relativistic particles. We study the angular smearing of ultrarelativistic charged leptons by the QED medium and the induced cross-section corrections in the lepton-nucleus interactions in Section~\ref{sec:electrons}, accounting for both initial- and final-state broadening. In the following Section~\ref{sec:neutrinos}, we evaluate the cross-section distortions in charged-current (anti)neutrino-nucleus reactions caused by multiple re-scattering of final-state charged leptons. We give our conclusions in Section~\ref{sec:conclusions} and present results for  different energy regimes in Appendices~\ref{app:electrons_other_energy} and~\ref{app:neutrinos_other_energy}. In Appendix~\ref{app:derivation}, we present diagrammatic calculation for cross-section corrections at the first three orders of the opacity expansion. Throughout the paper, we show results for several nuclei of interest to the neutrino and electron scattering communities.

\section{Multiple re-scattering in a QED nuclear medium}
\label{sec:formalism}

In this Section, we consider multiple QED re-scattering of charged leptons after the hard interaction inside the  nuclear medium. We provide estimates for the properties of such interactions and evaluate the transverse momentum distribution of the leptons leaving the nucleus.

After the hard scattering, the charged lepton propagates in the Coulomb field created by the electric charges inside the nucleus with a potential that in momentum space reads $v \left( \vec{q}_\perp \right) = \frac{4 \pi \alpha}{\vec{q}^2_\perp + \zeta^2}$, where $\zeta \approx \frac{m_e Z^{1/3}}{192}$ is regularization at the atomic screening scale, $Z$ is the nuclear charge number, $m_e$ is the electron mass, and $\alpha$ is the electromagnetic coupling constant. The total scattering cross section $\sigma_v$ of the lepton of mass $m_\ell$ with energy $E^\prime_\ell$ on a point source of this potential can be expressed as
\begin{align} 
\sigma_v \approx \frac{4 \pi \alpha^2}{\left( \frac{m_e Z^{1/3}}{192} \right)^2} \frac{1}{\beta_\ell^2},
\end{align}
with $\beta_\ell = \sqrt{ 1 - m^2_\ell/\left(E^\prime_\ell\right)^2}$, for sufficiently large $\beta_\ell \gg \alpha$ when the perturbation theory is still valid. Moving along its direction inside the nucleus, the lepton re-scatters on the nuclear charge distribution multiple times. The corresponding QED nuclear medium mean-free-path $\lambda_\mathrm{mfp}$ is expressed in terms of the nuclear size $R_\mathrm{rms}$ through the density of scattering centers $n \sim 1/R^3_\mathrm{rms}$ as
\begin{align} 
\lambda_\mathrm{mfp} = \frac{1}{n \sigma_v} = \frac{R_\mathrm{rms}}{3 Z^{1/3}} \left(\frac{\beta_\ell m_e R_\mathrm{rms}}{192 \alpha} \right)^2.
\end{align}
For numerical evaluations, we take nuclear radii from Ref.~\cite{nds_charge_radii}. To estimate the number of QED interactions $\chi$ inside the nucleus, we assume that the lepton travels the distance $R_\mathrm{rms}$ in the uniform charge distribution of the nucleus and divide the nuclear size by the mean-free-path length:
\begin{align}  \label{eq:number_of_scatterings}
\chi \sim \frac{R_\mathrm{rms}}{\lambda_\mathrm{mfp}} \sim  \frac{Z^{1/3}}{\left(\beta_\ell m_e R_\mathrm{rms} \right)^2}.
\end{align}
For ultrarelativistic leptons, i.e., $\beta_\ell \to 1$, the number of QED interactions reaches $\chi \sim 10^5-10^7$. This number increases for smaller lepton energies. In the relativistic regime that we consider $\beta_\ell \gg \alpha$, all re-scattering happens mainly in the forward direction, and the charged lepton can gain quite large transverse momentum $p^\prime_\perp$ relative to its direction of propagation with a resulting distribution~\cite{Ovanesyan:2011xy}
\begin{align} \label{eq:distribution_pT}
\frac{\mathrm{d} N}{\mathrm{d} p^\prime_\perp} =  \int \limits^{\infty}_{0} b p^\prime_\perp J_0 (0, b p^\prime_\perp) e^{- \chi \left( 1 - \zeta b K_1 \left( \zeta b \right) \right)} \mathrm{d} b,
\end{align}
expressed as an integral in the transverse coordinate space over the radial coordinate $b$, with the modified Bessel function of the second kind $K_1$ and the Bessel function of the first kind $J_0$. The $p^\prime_\perp$ distribution of Eq.~(\ref{eq:distribution_pT}) integrates to unity over all transverse momenta. We verified this expression explicitly at first $3$ orders in opacity in QED by proving that the additional diagrams, which vanish in QCD due to zero trace in the color space, do not contribute after summing over all diagrams and re-scattering sites. We sketch this cancellation in the Appendix~\ref{app:derivation}. The Gaussian approximation
\begin{align} \label{eq:distribution_pT_Gaussian}
\frac{\mathrm{d} N^G}{\mathrm{d} p^\prime_\perp} = \frac{p^\prime_\perp}{\chi \zeta^2 \xi} e^{-\frac{\left(p^\prime_\perp\right)^2}{2 \chi \zeta^2 \xi}},
\end{align}
with $\xi = \frac{1 - 2 \left[ \gamma_E - \ln 2 + \ln \left( \zeta R_0 \right)  \right]}{4}$ and nuclear scale $R_0 \sim 1~\mathrm{fm}$,  describes the transverse momentum distribution of Eq.~(\ref{eq:distribution_pT}) relatively well.

\begin{figure}[!t]
    \centering
    \includegraphics[height=0.84\textwidth]{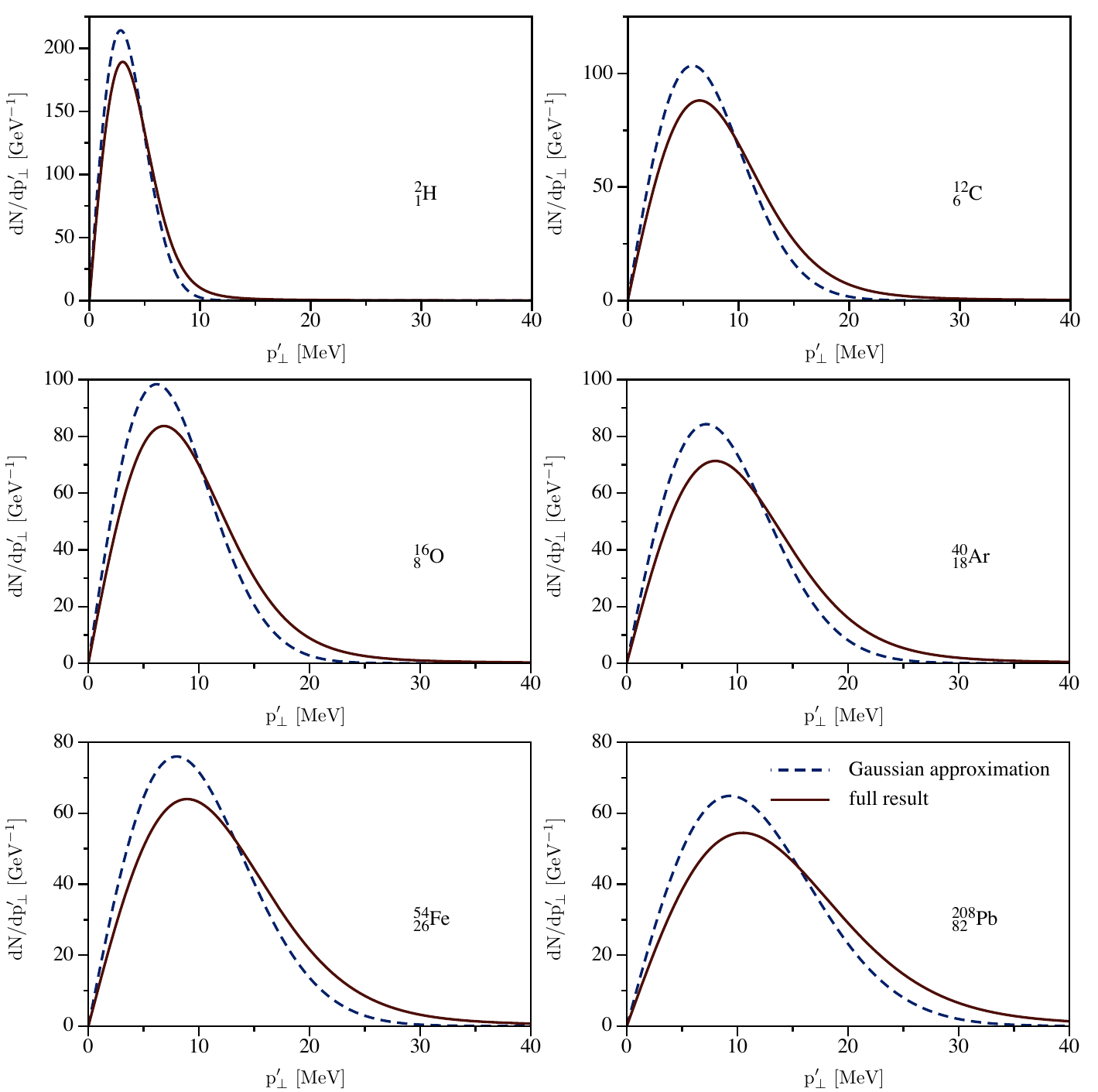}
    \caption{The transverse momentum distribution of the final-state ultrarelativistic charged lepton escaping the QED nuclear medium relative to the original final-state direction is presented for various nuclei of interest to neutrino and electron scattering communities. The broadening is assumed to happen on the scale of the nuclear radius.}\label{fig:broadening}
\end{figure}

In the following Fig.~\ref{fig:broadening}, we present the transverse momentum distribution both for the numerical integration of Eq.~(\ref{eq:distribution_pT}), which we use in the following chapters of this paper, and for the Gaussian approximation of Eq.~(\ref{eq:distribution_pT_Gaussian}). Results are shown for ultrarelativistic electrons and typical in neutrino and electron scattering experiments nuclei $^2_1\mathrm{H}$, $^{12}_6\mathrm{C}$, $^{16}_8\mathrm{O}$, $^{40}_{18}\mathrm{Ar}$, $^{56}_{26}\mathrm{Fe}$, and $^{208}_{82}\mathrm{Pb}$. However, these results apply also for the ultrarelativistic muons since only the parameter $\beta_\ell$ distinguishes the lepton flavor. The strength of the $p^\prime_\perp$ peak decreases while the average $p^\prime_\perp$ increases with the nuclear size. The Gaussian approximation results in $10$-$20~\%$ lower peak strength and suppresses the high-momentum tails compared to the exact result for the transverse momentum distribution. Considering the $2\sigma$ range of the Gaussian distribution, the charged lepton gains around $8$-$26~\mathrm{MeV}$ of transverse momentum while traveling in the nuclear medium. The size of the acquired transverse momentum is above the typical experimental resolution. Therefore, the broadening of charged lepton trajectories is a measurable effect that should be included in the analysis of precise electron- and muon-nucleus scattering experiments.

\section{Broadening of electrons by the nuclear medium} \label{sec:electrons}

In this Section, we calculate the deflection of charged lepton trajectories inside the nucleus, accounting for the multiple QED re-scattering before and after the hard interaction, and investigate the induced changes in measured cross sections.

First, we consider the charged lepton re-scattering after the hard interaction. QED nuclear broadening does not change the lepton energy but redistributes the momentum from the initial direction by contributing to the transverse direction in an azimuthally-symmetric manner. Let us denote the lepton scattering angle in the hard process by $\theta^0_\ell$ and select the coordinate frame such that the $3$-momentum of the lepton before the hard interaction is along the $z$ axis. After the hard interaction, the $3$-momentum becomes $\beta_\ell E^\prime_\ell \left( \sin \theta^0_\ell, 0, \cos \theta^0_\ell \right)$. Leaving the nucleus, the lepton gets a kick by soft QED re-scattering with a momentum $p^\prime_\perp$ in the orthogonal direction $\left(p^\prime_L \sin \theta^0_\ell, p^\prime_\perp, p^\prime_L \cos \theta^0_\ell \right)$, where $p^\prime_L = \sqrt{ \left( E^\prime_\ell \right)^2 - \left( p^\prime_\perp \right)^2 - m_\ell^2}$, and the direction of $p^\prime_\perp$ is taken along the $y$ axis, for definiteness. For the perpendicular component with a polar angle $\phi$, the outgoing lepton scattering angle $\theta_\ell$ w.r.t. to the direction right before the hard interaction is determined as
\begin{align} \label{eq:lepton_angle}
    \cos \theta_\ell = \frac{p^\prime_L \cos \theta^0_\ell + p^\prime_\perp \sin \theta^0_\ell \sin \phi}{\sqrt{ \left( p^\prime_\perp \right)^2 + \left( p^\prime_L \right)^2}}.
\end{align}

Let us consider the elastic scattering of electrons on nucleons inside the nucleus. Since broadening does not change the lepton energy, there is no change in the cross section when the momentum transfer $Q^2$ in the elastic process on nucleons is defined from the initial ($E_\mathrm{beam}$) and final ($E^\prime_\ell$) electron energies as $Q^2 = 2 M \left( E_\mathrm{beam} - E^\prime_\ell \right)$. In our calculation, we consider the scattering off nucleons at rest and do not include radiative corrections to the hard process. In elastic electron-nucleus reactions, the multiple re-scattering will redistribute the incoming and outgoing electron angles compared to the kinematical value that corresponds to the final electron energy. Accounting for the nuclear medium effects on the final-state electron, the $r.m.s.$ scattering angle distortion $\sqrt{\left<\left(\Delta \theta \right)^2 \right>}$ can be expressed as
\begin{align}
\sqrt{\left<\left(\Delta \theta \right)^2 \right>} = \sqrt{\int  \left( \theta_\ell - \theta^0_\ell \right)^2 \frac{\mathrm{d} N}{\mathrm{d} p^\prime_\perp} \mathrm{d} p^\prime_\perp \frac{\mathrm{d} \phi}{2\pi}}.
\end{align}
We also include the deflection of the initial-electron angle and average over the interaction sites inside the nucleus with a corresponding change of the re-scattering length in Eq.~(\ref{eq:number_of_scatterings}), which enters the evaluation of the transverse momentum distribution $\mathrm{d} N / \mathrm{d} p_\perp$, before and after the hard interaction. For this calculation, we consider the nucleus as a sphere of the nuclear radius $R_\mathrm{rms}$ with a uniformly distributed charge and assume broadening angles to be small. In the following Fig.~\ref{fig:electron_broadening_2}, we present the $r.m.s.$ deflection angle $\Delta \theta \equiv \sqrt{\left<\left(\Delta \theta \right)^2 \right>}$ w.r.t. to the direction determined by the elastic kinematics as a function of the recoil electron energy $E^\prime_e$ for the beam energy $E_\mathrm{beam} = 2~\mathrm{GeV}$. The $r.m.s.$ deflection angle reaches $1$-$1.5^\mathrm{o}$ and decreases with increasing the recoil electron energy. As expected, the broadening increases with the size of the nucleus. Such a large effect is above the typical resolution of experiments at $\mathrm{GeV}$ energies and can be measured in the electron-nucleus scattering by considering the angular distributions for a one-nucleon knockout reaction for the fixed recoil electron energy. However, radiative events should be carefully disentangled from the measured distribution. Our results apply also to the elastic muon-nucleus scattering up to power-suppressed corrections of order $m^2_\mu/E^2_\mu$.
\begin{figure}[ht]
    \centering
      \includegraphics[height=0.33\textwidth]{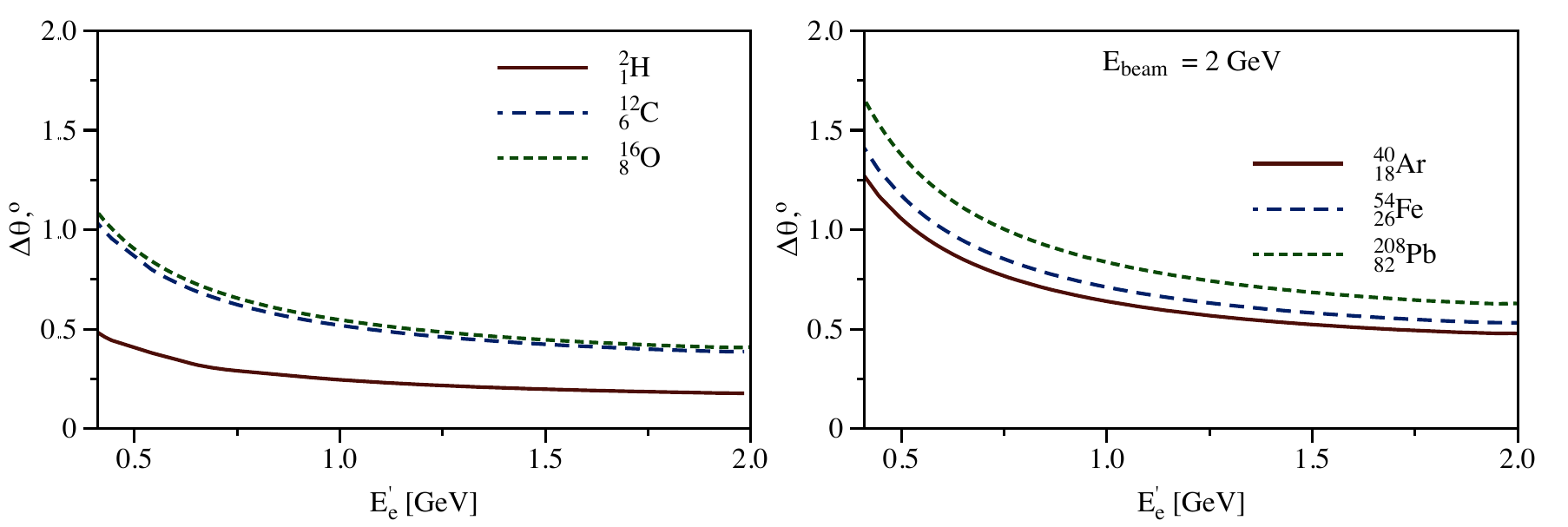}
    \caption{The $r.m.s.$ broadening angle is shown as a function of the recoil electron energy $E^\prime_e$ in the scattering of $2~\mathrm{GeV}$ electrons off nuclear targets. Both initial and final re-scattering are taken into account.}\label{fig:electron_broadening_2}
\end{figure}

\begin{figure}[ht]
    \centering
      \includegraphics[height=0.33\textwidth]{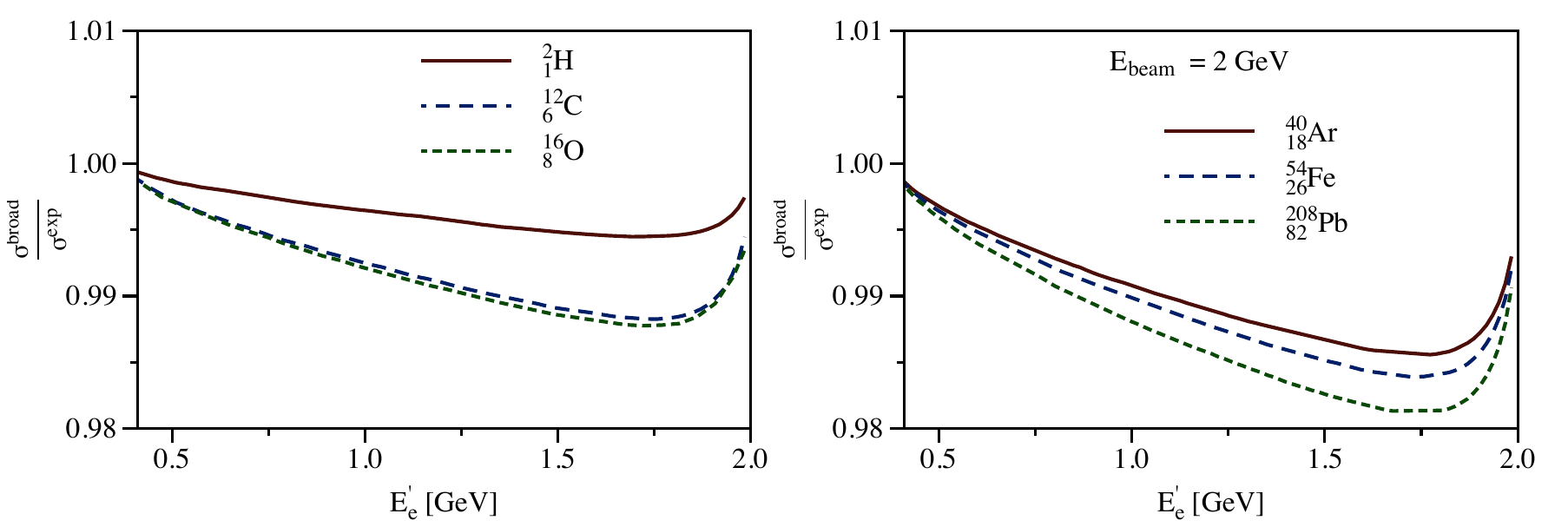}
    \caption{Ratio of the elastic electron-proton scattering cross section after accounting for the QED broadening of electron tracks inside the nucleus to the cross section predicted from the lepton kinematics as a function of the recoil electron energy $E^\prime_e$ in the scattering of $2~\mathrm{GeV}$ electrons off nuclear targets. Both initial and final re-scattering are taken into account.}\label{fig:electron_xsection_2}
\end{figure}

Angular smearing can modify experimentally measured single-nucleon cross sections when the momentum transfer is reconstructed from the initial- and final-state lepton kinematics. For the fixed by the elastic kinematics scattering angle and recoil energy, hard processes with different momentum transfers contribute to the cross section. We consider the ratio of averaged over scattering angles of the hard process ``true" cross section $\sigma^\mathrm{broad}$ to the expected cross section $\sigma^\mathrm{exp}$, which we evaluate from the lepton kinematics with elastic condition between the recoil lepton energy and scattering angle. We present the corresponding ratio for electron-proton scattering with the electron beam energy $E_\mathrm{beam} = 2~\mathrm{GeV}$ in Fig.~\ref{fig:electron_xsection_2}. Similarly to the broadening angle, the size of cross-section suppression increases with the size of the nucleus. Surprisingly, we find a new effect that can reach $1$-$2\%$ for electron scattering off heavy nuclei. Such large corrections should be considered in modern and future analyses of the electron scattering data. The size of broadening decreases with the beam energy approximately as $\sqrt{\left<\left(\Delta \theta \right)^2 \right>} \sim {1}/{E}$. We present results for other beam energies in Appendix~\ref{app:electrons_other_energy}.

\section{Charged lepton broadening in (anti)neutrino-nucleus scattering} \label{sec:neutrinos}

In this Section, we quantify the effects of the charged lepton broadening on the (anti)neutrino-nucleon $\nu_\ell n \to \ell^- p$ and $\bar{\nu}_\ell p \to \ell^+ n$ cross sections inside the QED nuclear medium.

Contrary to electron and muon scattering experiments, the incoming (anti)neutrino flux is broad in energy and the initial (anti)neutrino energy has to be determined experimentally. The reconstructed (anti)neutrino energy $E_\nu^r$ is evaluated from the final lepton energy $E^\prime_\ell$ and the scattering angle $\theta_\ell$ as~\cite{Bodek:2007wb}
\begin{align}
    E_\nu^r = \frac{E^\prime_\ell- \frac{1}{2} \frac{E_B^2 -2 M_1 E_B + m_\ell^2 + M^2_1 - M^2_2}{M_1-E_B}}{1- \frac{E^\prime_\ell}{M_1-E_B}  \left( 1 - \beta_\ell \cos \theta_\ell \right)},
\end{align}
with the masses of initial and final nucleons $M_1$ and $M_2$, respectively, and the binding energy $E_B$. The binding energy for $^{40} \mathrm{Ar}$, is of order $10~\mathrm{MeV}$: $E_B^n = 9.869~\mathrm{MeV}$ for neutrons and $E_B^p = 12.5268~\mathrm{MeV}$ for protons. We take the binding energies from the AME 2020 atomic mass evaluation tables~\cite{Wang:2012eof,Wang:2021xhn}. Having determined the incoming (anti)neutrino energy, we assign the momentum transfer $Q^2$ for each event as
\begin{align}
Q^2 = -m_\ell^2 +2 E^r_\nu E^\prime_\ell \left( 1-\beta_\ell \cos \theta_\ell \right).
\end{align}
\begin{figure}[ht]
    \vspace{-1.cm}
    \centering
    \includegraphics[height=0.33\textwidth]{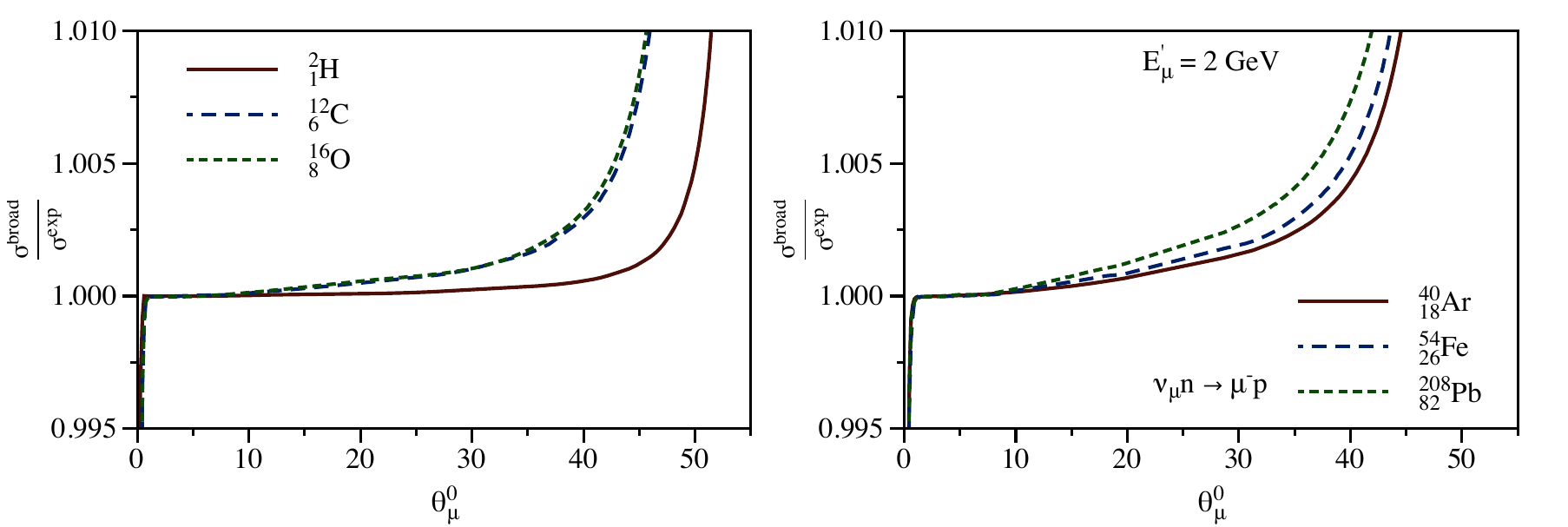}
    \caption{Ratio of the charged-current elastic muon neutrino-neutron cross section after accounting for the broadening to the cross section predicted from the lepton kinematics as a function of the measured angle for outgoing muon with the energy $E^\prime_\mu = 2~\mathrm{GeV}$.}\label{fig:neutrino_broadening_2}
    \centering
    \includegraphics[height=0.33\textwidth]{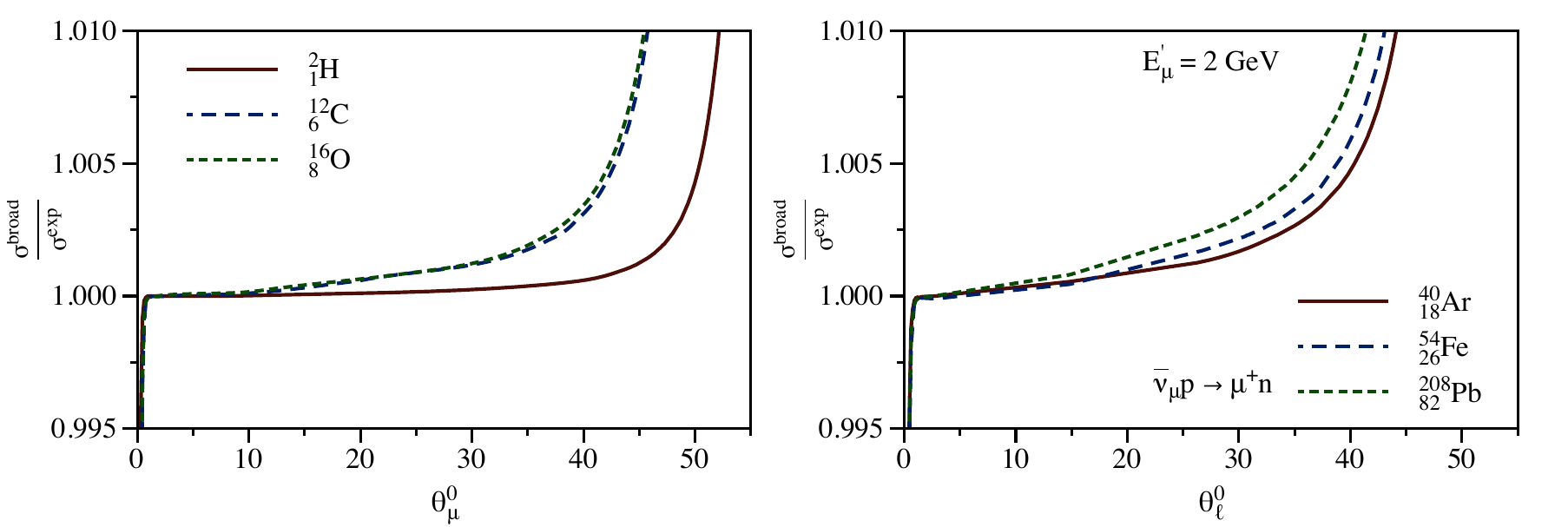}
    \caption{Same as Fig.~\ref{fig:neutrino_broadening_2}, but for the antineutrino scattering.}\label{fig:antineutrino_broadening_2}
\end{figure}

The difference in the measured azimuthal angle and the ``true" scattering angle in the hard process modifies the experimentally extracted single-nucleon cross section such that the ratio of the measured cross section $\sigma^\mathrm{broad}$ to the predicted from the lepton kinematics expectation $\sigma^\mathrm{exp}$ is given by
\begin{align} \label{eq:ratio_broadening_neutrino}
  \frac{\sigma^\mathrm{broad}}{\sigma^\mathrm{exp}} = \int \frac{\frac{\mathrm{d} \sigma \left(E^r_\nu \left(\theta_\ell \right), \theta_\ell \right)}{\mathrm{d} Q^2}}{\frac{\mathrm{d} \sigma \left(E^r_\nu \left(\theta^0_\ell \right), \theta^0_\ell \right)}{\mathrm{d} Q^2}} \frac{\mathrm{d} N}{\mathrm{d} p^\prime_\perp} \mathrm{d} p^\prime_\perp \frac{\mathrm{d} \phi}{2\pi} \delta \left( \cos \theta_\ell - \frac{p^\prime_L \cos \theta^0_\ell + p^\prime_\perp \sin \theta^0_\ell \sin \phi}{\sqrt{\left(p^\prime_\perp \right)^2 + \left( p^\prime_L \right)^2}} \right) \mathrm{d} \cos \theta_\ell,
\end{align}
where we take just one solution for the angle $\theta^0_\ell$ and exploit the expression for the lepton azimuthal angle from Eq.~(\ref{eq:lepton_angle}). As in Section~\ref{sec:electrons}, we consider the nucleus as a sphere of the nuclear radius $R_\mathrm{rms}$ with a uniformly distributed charge. In the following Figs.~\ref{fig:neutrino_broadening_2}-\ref{fig:antineutrino_broadening_2}, we show the ratio defined in Eq.~(\ref{eq:ratio_broadening_neutrino}) for muon neutrino and antineutrino scattering off neutrons and protons, respectively.

The relative cross-section corrections are typically below or at the permille level. In the forward direction, we observe a sizable suppression of cross sections, which can be large close to the lepton production threshold. At the same time,  enhancement is observed for the largest (allowed by the kinematics) scattering angles
\begin{align} \label{eq:extreme_angles}
    \cos \theta_\ell^\mathrm{max} \approx \frac{1}{\beta_\ell} - \frac{M_1 - E_B}{\beta_\ell E^\prime_\ell},
\end{align}
when the recoil lepton energy $E^\prime_\ell > \frac{M_1 - E_B}{2} + \frac{1}{2} \frac{m^2_\ell}{M_1 - E_B}$. The corresponding corrections can exceed the percent level. The scattering into the largest angles defined in Eq.~(\ref{eq:extreme_angles}) is realized only for (anti)neutrinos of high energy, while the forward angles are reached  nearby the lepton production threshold. Performing the averaging over energies of the typical neutrino flux~\cite{dune_page}, we obtain the corresponding cross-section corrections a way below the permille level since the broadening of lepton trajectories only re-distributes the scattering events.

\section{Conclusions}
\label{sec:conclusions}

In this paper, we investigated the effects of multiple re-scattering of charged leptons mediated by QED interactions inside a nuclear medium. We found the typical induced transverse momentum of outgoing ultrarelativistic leptons at the level $10$-$30~\mathrm{MeV}$ depending on the nucleus under consideration. The broadening magnitude increases with the size of the nucleus. At hundred of MeV to GeV energies, the corresponding QED nuclear medium re-scattering deflects the trajectories of relativistic charged leptons by a few degrees. The $r.m.s.$ deflection angle decreases with the corresponding energy scale $E$ as $\sqrt{\left<\left(\Delta \theta \right)^2 \right>} \sim {1}/{E}$. Contrary to our previous unresummed results, the corresponding cross-section modifications in electron-nucleus scattering can reach few-percent level over wide range of the kinematics and not only in the forward direction. It would be important to measure such effects and unfold them in the extraction of nucleon and nuclear structure with electron-nucleus scattering experiments, when the incoming lepton energy is known. In (anti)neutrino scattering experiments with broad in energy incoming flux and unknown initial-lepton energy, multiple QED re-scattering inside the nucleus modifies the experimentally-accessed cross sections typically at permille level and below. Only near the kinematic endpoints, cross-section distortions can reach a percent level with a suppression in the forward direction and enhancement for large-angle scattering.

We note that in addition to tree-level scattering effects, interactions in the QED nuclear medium can induce radiative corrections to the scattering cross sections. In the case of QCD, they were initially interpreted  as energy loss of the strongly-interacting quarks and gluons~\cite{Baier:2000mf,Gyulassy:2000er,Wang:2001ifa,Arnold:2002ja}. In electron-nucleus reactions in the deep inelastic scattering regime, radiative processes can significantly suppress the cross sections for semi-inclusive hadron and jet production~\cite{Chang:2014fba,Li:2020zbk,Li:2021gjw,Li:2020rqj,Ke:2023xeo,Li:2023dhb}. Interactions in QCD matter further induce photon bremsstrahlung~\cite{Arnold:2002ja,Zakharov:2004bi,Vitev:2008vk}. The corresponding radiative corrections from QED interactions in large nuclei have not been considered before and we leave them for future work.

\section*{Acknowledgments}

The work is supported by the US Department of Energy through the Los Alamos National Laboratory. Los Alamos National Laboratory is operated by Triad National Security, LLC, for the National Nuclear Security Administration of the U.S. Department of Energy (Contract No. 89233218CNA000001). This research is funded by LANL’s Laboratory Directed Research and Development (LDRD/PRD) program under project numbers 20210968PRD4 and 20240127ER. FeynCalc~\cite{Mertig:1990an,Shtabovenko:2016sxi}, LoopTools~\cite{Hahn:1998yk}, Mathematica~\cite{Mathematica}, and DataGraph~\cite{JSSv047s02} were used in this work.

\appendix

\section{Electron broadening at different beam energies} \label{app:electrons_other_energy}

\begin{figure}[ht]
    \vspace{1.0cm}
    \centering
      \includegraphics[height=0.33\textwidth]{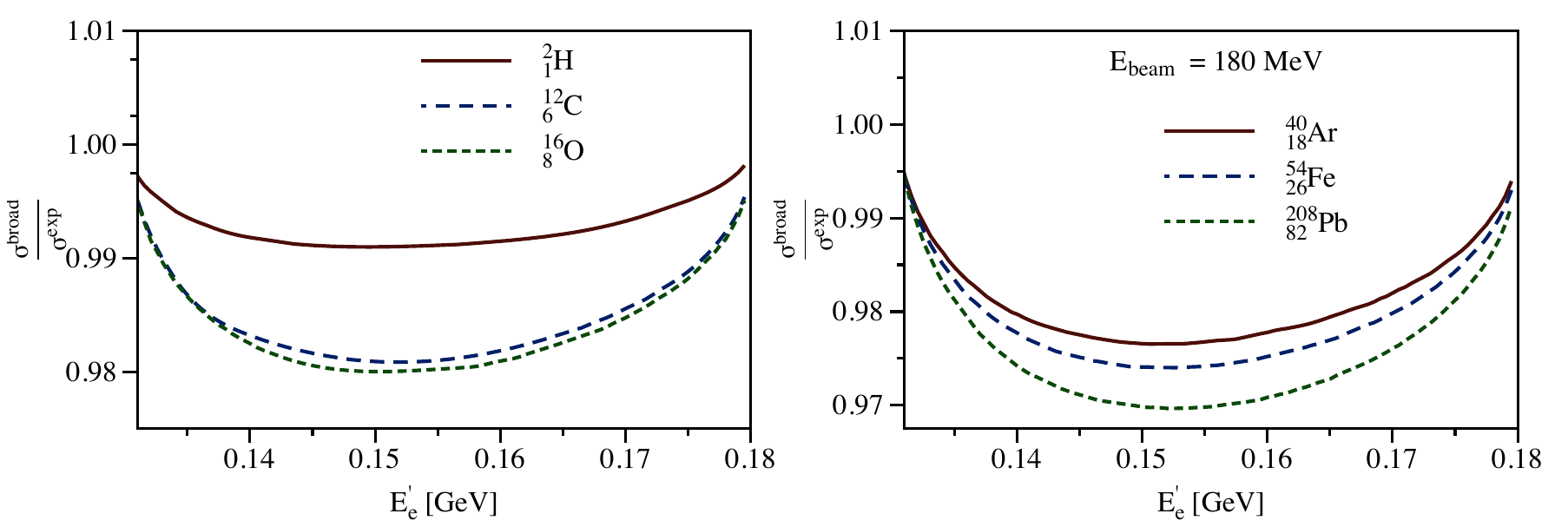}
    \caption{Same as Fig.~\ref{fig:electron_xsection_2}, but for the electron beam energy $E_\mathrm{beam} = 180~\mathrm{MeV}$.} \label{fig:electron_xsection_180}
\end{figure}

\begin{figure}[ht]
    \vspace{1.0cm}
    \centering
      \includegraphics[height=0.33\textwidth]{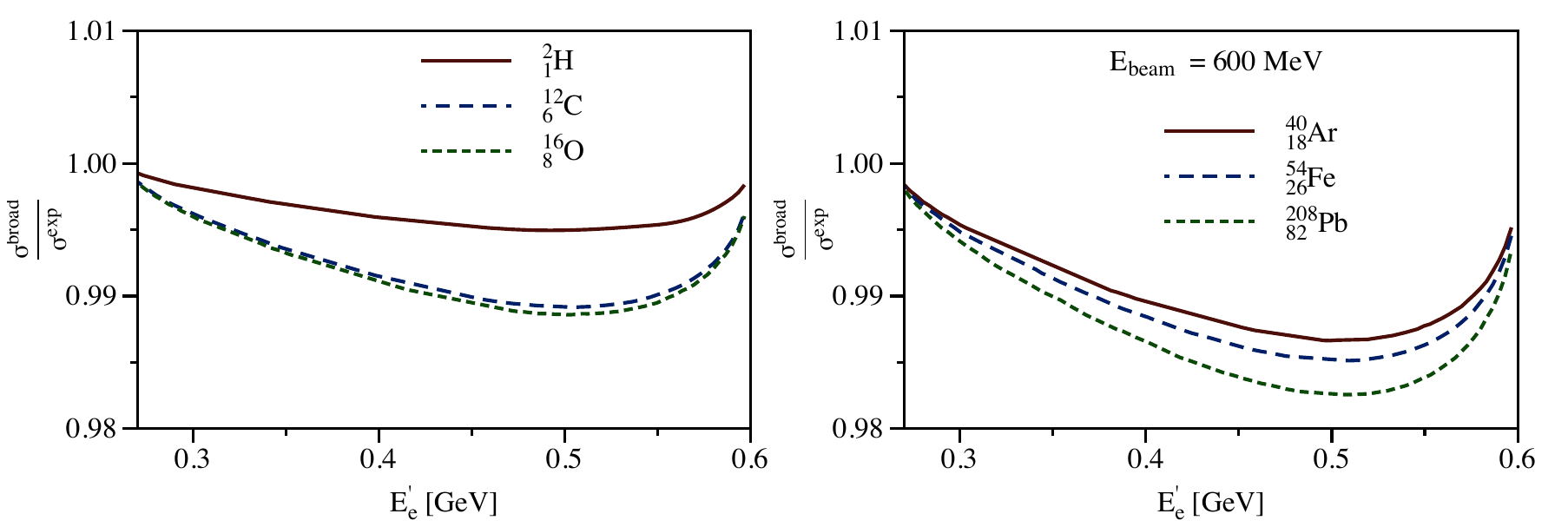}
    \caption{Same as Fig.~\ref{fig:electron_xsection_2}, but for the electron beam energy $E_\mathrm{beam} = 600~\mathrm{MeV}$.} \label{fig:electron_xsection_600}
\end{figure}

\begin{figure}[ht]
    \vspace{1.0cm}
    \centering
     \includegraphics[height=0.33\textwidth]{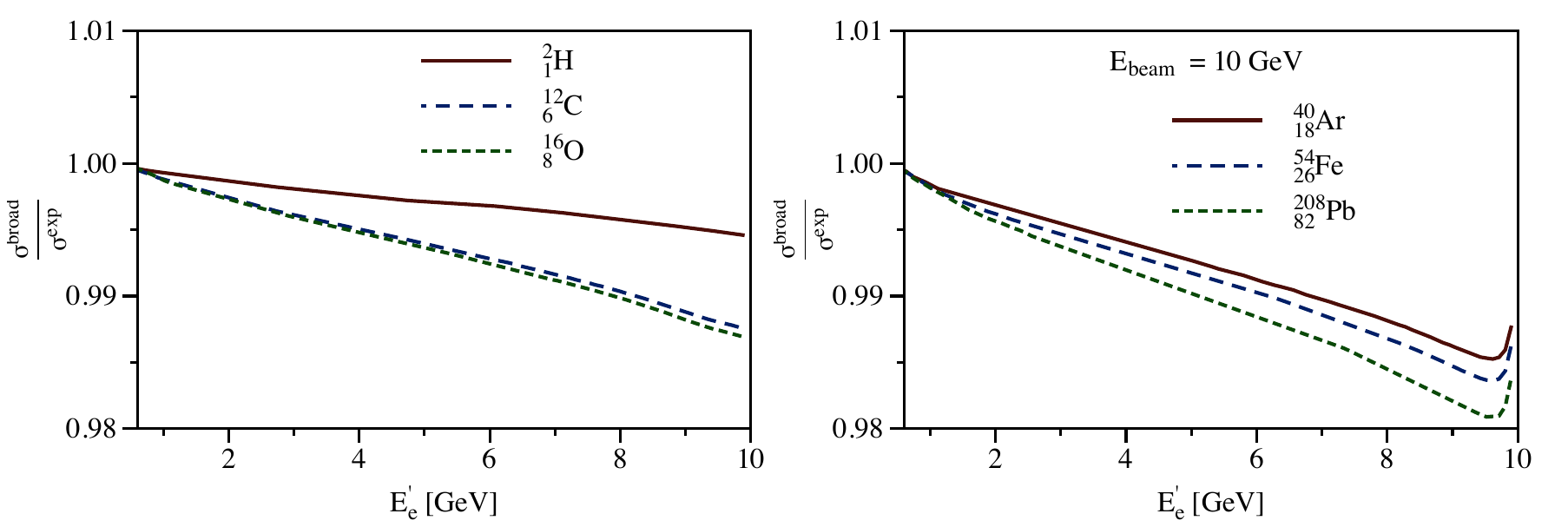}
    \caption{Same as Fig.~\ref{fig:electron_xsection_2} but for the electron beam energy $E_\mathrm{beam} = 10~\mathrm{GeV}$.} \label{fig:electron_xsection_10}
\end{figure}

In this Appendix, we present the $r.m.s.$ deflection angle in the electron-nucleus scattering after averaging over the nuclear size for the lowest beam energy of electron scattering experiments at MAMI/Mainz $E_\mathrm{beam} = 180~\mathrm{MeV}$~\cite{Bernauer:2010zga,A1:2010nsl,A1:2013fsc}, for the peak of T2K/HyperK experiment $E_\mathrm{beam} = 600~\mathrm{MeV}$~\cite{T2K:2011qtm,T2K:2019bcf,Hyper-KamiokandeProto-:2015xww}, and for the higher energy $E_\mathrm{beam} = 10~\mathrm{GeV}$ in the following Figs.~\ref{fig:electron_broadening_180}-\ref{fig:electron_broadening_10}. As noted in Ref.~\cite{Tomalak:2022kjd}, the two lowest energies correspond to the extrapolation of the effective field theory calculation. The broadening of the electron trajectories increases at lower energies. The dependence on the electron beam energy can be roughly approximated as $\sqrt{\left<\left(\Delta \theta \right)^2 \right>} \sim {1}/{E_\mathrm{beam}}$ with large deflection angles at hundreds MeV energy. We present the corresponding ratio of the elastic electron-proton scattering cross section after accounting for the QED broadening of electron tracks inside the nucleus to the cross section predicted from the lepton kinematics for energies $E_\mathrm{beam} = 180~\mathrm{MeV}$, $E_\mathrm{beam} = 600~\mathrm{MeV}$, and $E_\mathrm{beam} = 10~\mathrm{GeV}$ in Figs.~\ref{fig:electron_xsection_180}-\ref{fig:electron_xsection_10}. The broadening effect increases over all scattering angles going to lower energies, while the largest by magnitude correction remains of the similar size. Increasing the electron beam energy, the position of largest suppression shifts to forward scattering with low values of the momentum transfer, where the slope of the cross section with respect to a recoil energy is significantly enhanced compared to the broadening redistribution.
\begin{figure}[ht]
    \vspace{1.0cm}
    \centering
      \includegraphics[height=0.33\textwidth]{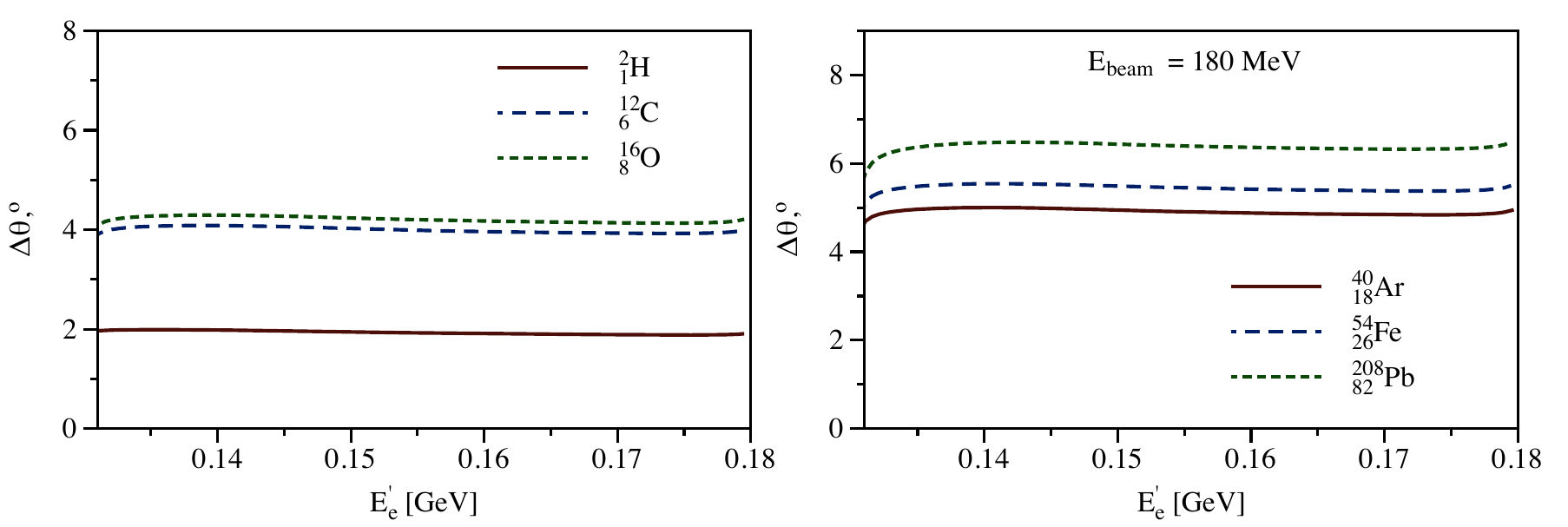}
    \caption{Same as Fig.~\ref{fig:electron_broadening_2}, but for the electron beam energy $E_\mathrm{beam} = 180~\mathrm{MeV}$.} \label{fig:electron_broadening_180}
\end{figure}
\begin{figure}[ht]
    \vspace{1.0cm}
    \centering
      \includegraphics[height=0.33\textwidth]{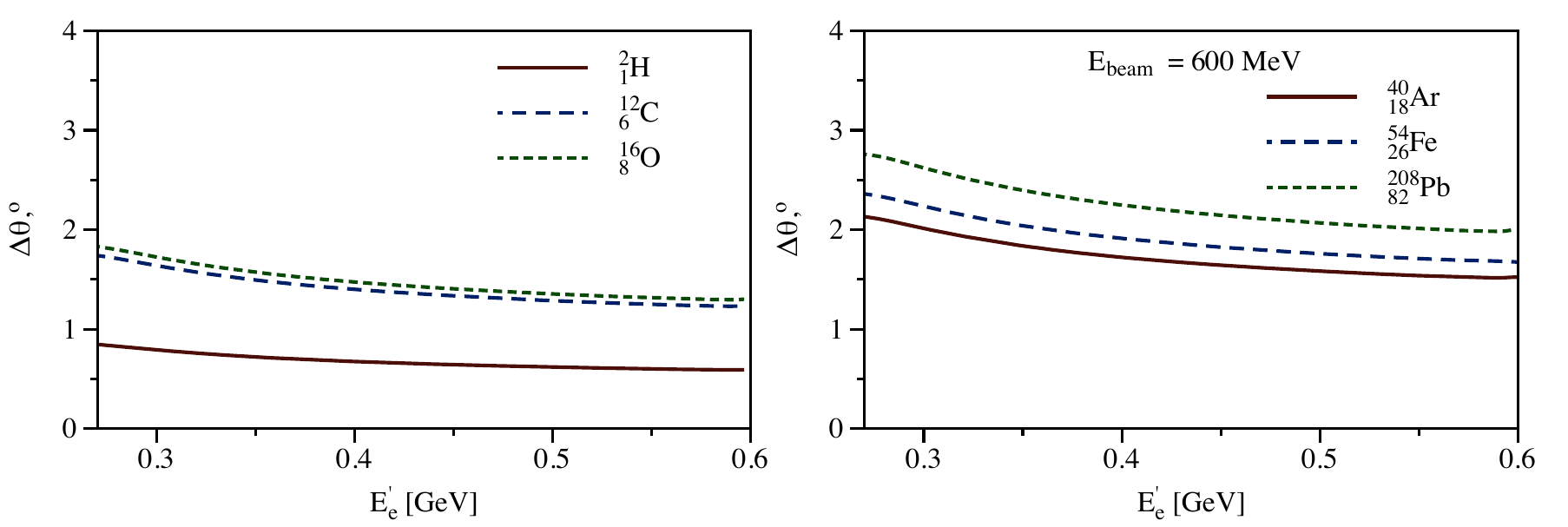}
    \caption{Same as Fig.~\ref{fig:electron_broadening_2}, but for the electron beam energy $E_\mathrm{beam} = 600~\mathrm{MeV}$.} \label{fig:electron_broadening_600}
\end{figure}
\begin{figure}[ht]
    \vspace{1.0cm}
    \centering
     \includegraphics[height=0.33\textwidth]{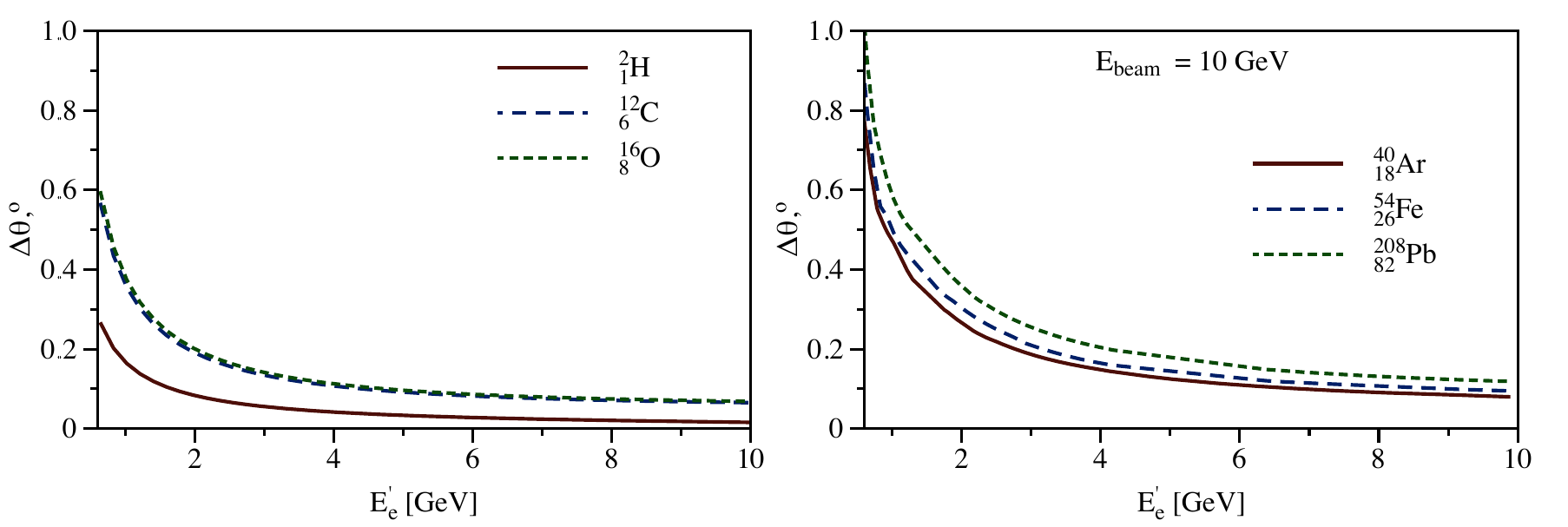}
    \caption{Same as Fig.~\ref{fig:electron_broadening_2} but for the electron beam energy $E_\mathrm{beam} = 10~\mathrm{GeV}$.} \label{fig:electron_broadening_10}
\end{figure}

\section{(Anti)neutrino-nucleus scattering at different beam energies\label{app:neutrinos_other_energy}}

\begin{figure}[]
    \centering
    \includegraphics[height=0.33\textwidth]{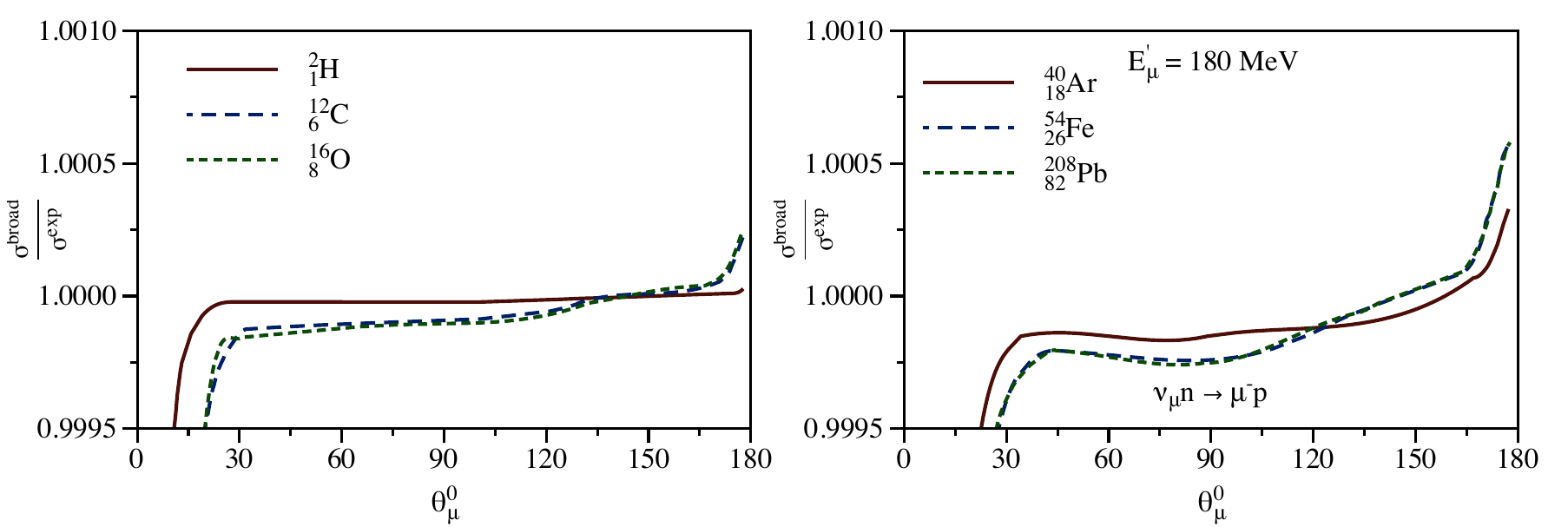}
    \caption{Same as Fig.~\ref{fig:neutrino_broadening_2} but for the outgoing muon energy $E^\prime_\mu = 180~\mathrm{MeV}$.} \label{fig:neutrino_broadening_180}
\end{figure}

\begin{figure}[]
    \centering
    \includegraphics[height=0.33\textwidth]{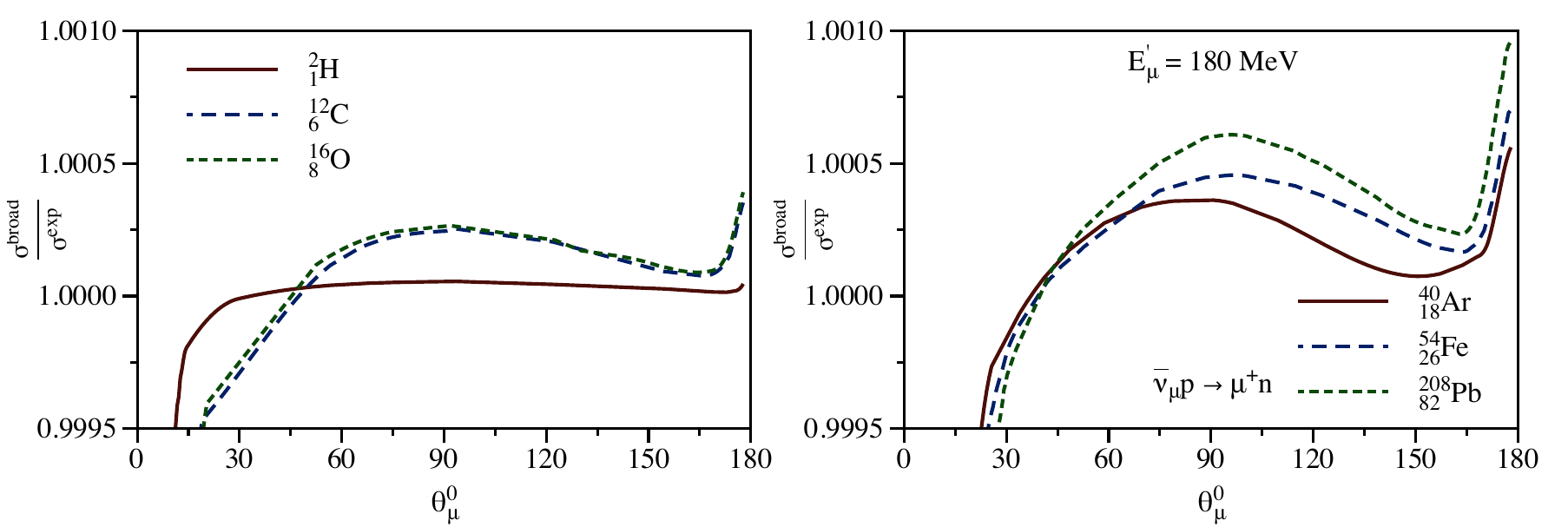}
    \caption{Same as Fig.~\ref{fig:antineutrino_broadening_2} but for the outgoing antimuon energy $E^\prime_\mu = 180~\mathrm{MeV}$.} \label{fig:antineutrino_broadening_180}
\end{figure}

\begin{figure}[]
    \centering
    \includegraphics[height=0.33\textwidth]{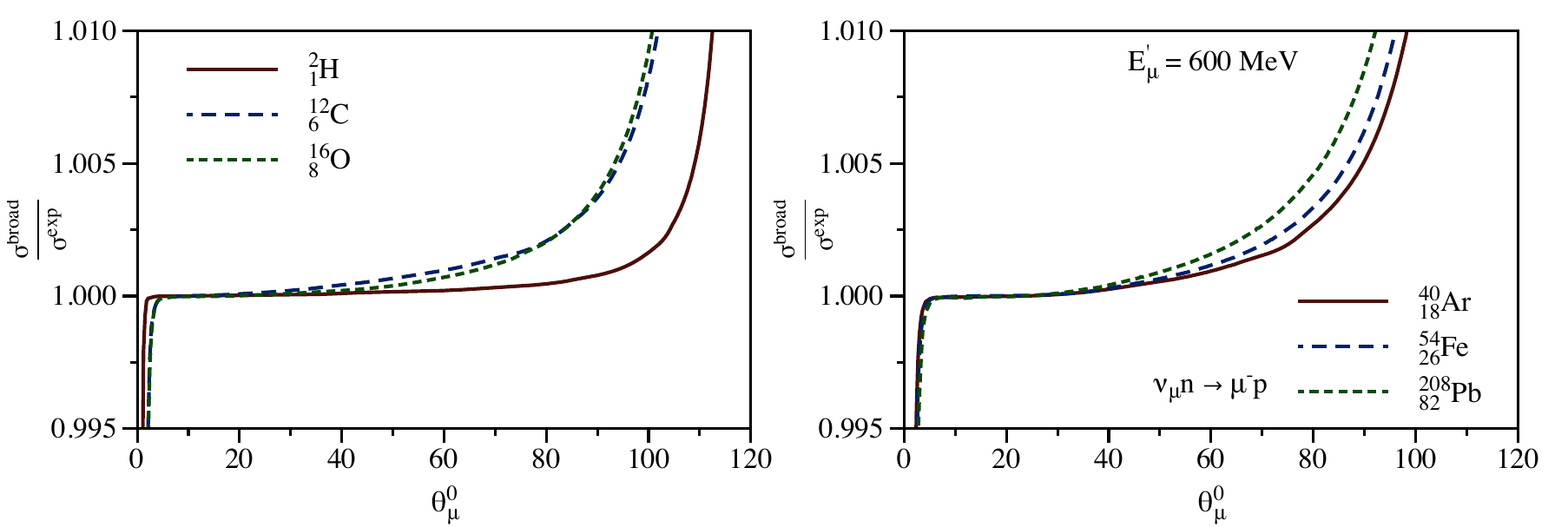}
    \caption{Same as Fig.~\ref{fig:neutrino_broadening_2} but for the outgoing muon energy $E^\prime_\mu = 600~\mathrm{MeV}$.}\label{fig:neutrino_broadening_600}
\end{figure}

\begin{figure}[]
    \centering
    \includegraphics[height=0.33\textwidth]{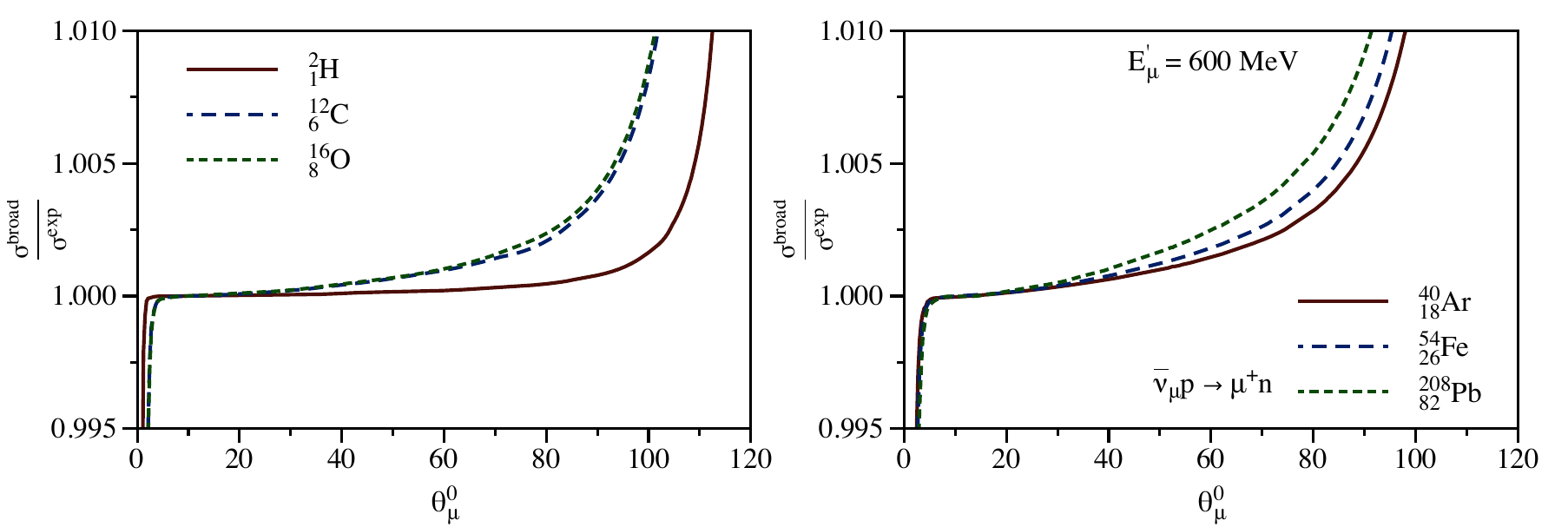}
    \caption{Same as Fig.~\ref{fig:antineutrino_broadening_2} but for the outgoing antimuon energy $E^\prime_\mu = 600~\mathrm{MeV}$.}\label{fig:antineutrino_broadening_600}
\end{figure}

\begin{figure}[]
    \centering
    \includegraphics[height=0.33\textwidth]{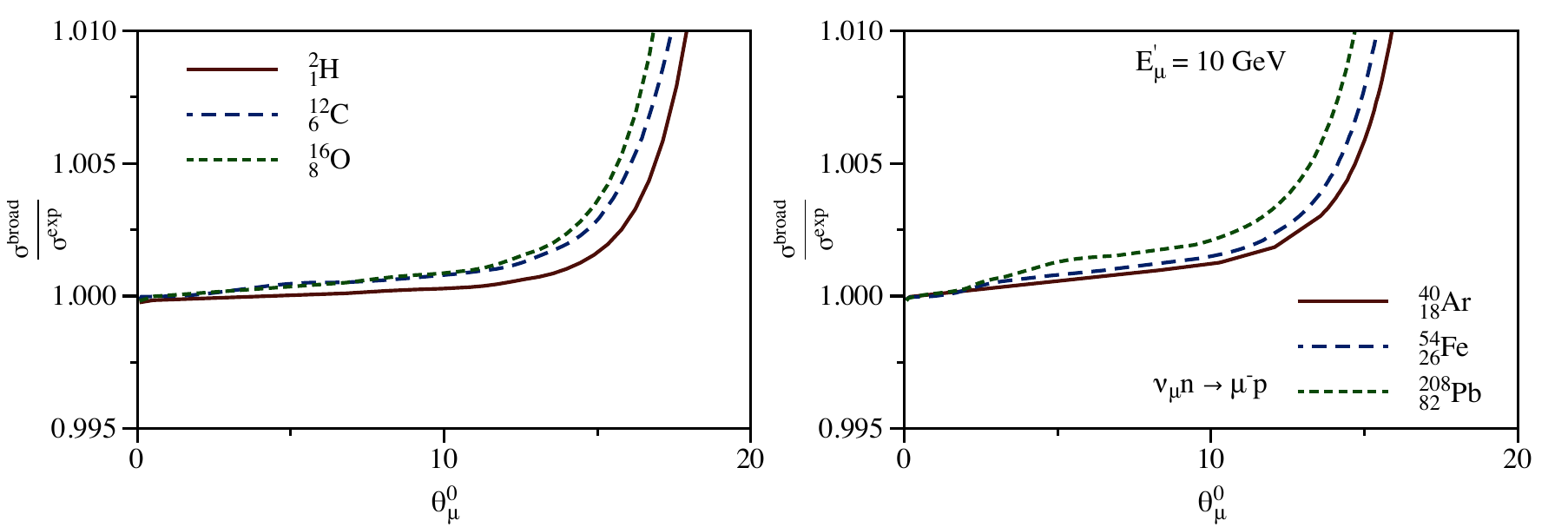}
    \caption{Same as Fig.~\ref{fig:neutrino_broadening_2} but for the outgoing muon energy $E^\prime_\mu = 10~\mathrm{GeV}$.}\label{fig:neutrino_broadening_10}
\end{figure}

\begin{figure}[]
    \centering
    \includegraphics[height=0.33\textwidth]{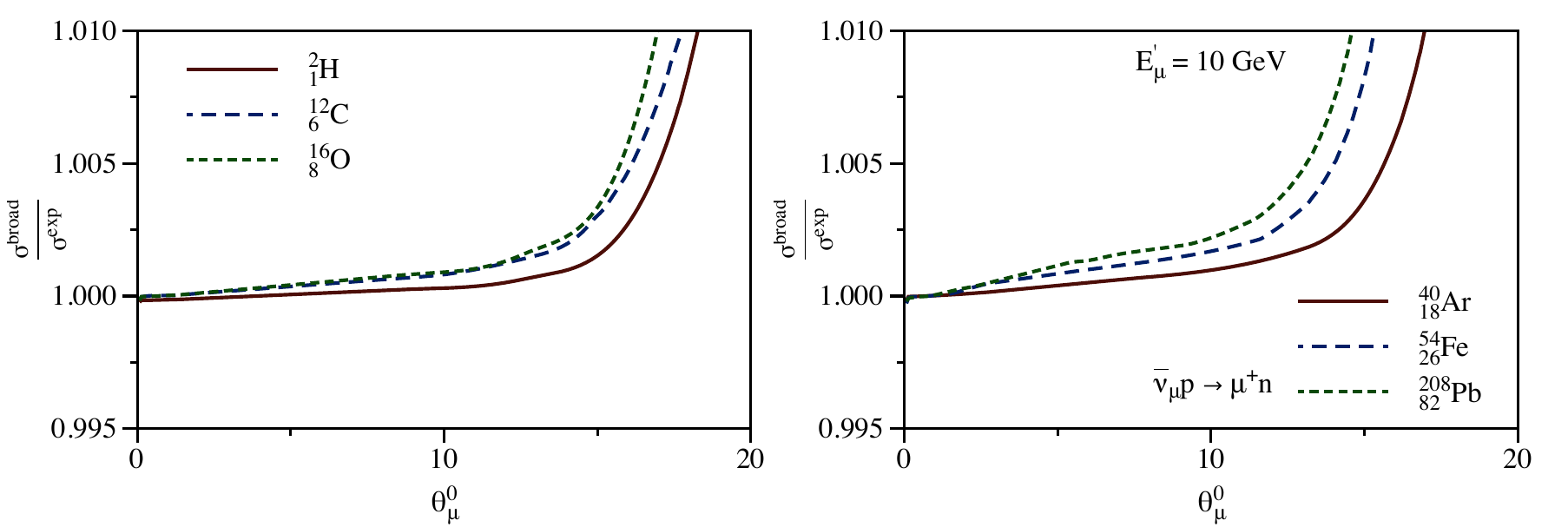}
    \caption{Same as Fig.~\ref{fig:antineutrino_broadening_2} but for the outgoing antimuon energy $E^\prime_\mu = 10~\mathrm{GeV}$.}\label{fig:antineutrino_broadening_10}
\end{figure}

In this Appendix, we present the results for the ratio of the cross section after broadening to the naive baseline cross section, cf. Eq.~(\ref{eq:ratio_broadening_neutrino}), as a function of the lepton scattering angle for recoil charged lepton energies $E^\prime_\ell = 180~\mathrm{MeV}$, corresponding to electron scattering experiments at MAMI/Mainz~\cite{Bernauer:2010zga,A1:2013fsc} and close to the peak of the future (anti)neutrino experiments with kaon-decay-at-rest sources~\cite{Spitz:2012gp,MiniBooNE:2018dus}, in Figs.~\ref{fig:neutrino_broadening_180}-\ref{fig:antineutrino_broadening_180}, and for the peak of T2K/HyperK experiment $E^\prime_\ell = 600~\mathrm{MeV}$~\cite{T2K:2011qtm,T2K:2019bcf,Hyper-KamiokandeProto-:2015xww} in Figs.~\ref{fig:neutrino_broadening_600}-\ref{fig:antineutrino_broadening_600}. We further show the results at higher energy $E^\prime_\ell = 10~\mathrm{GeV}$ in Figs.~\ref{fig:neutrino_broadening_10}-\ref{fig:antineutrino_broadening_10}. As noted in Ref.~\cite{Tomalak:2022kjd}, the two lowest energies correspond to the extrapolation of the effective field theory calculation. For the recoil lepton energy $E^\prime_\ell = 180~\mathrm{MeV}$, all scattering angles are allowed by the kinematics. At different energies, we observe the same pattern of increasing with a nuclear size cross-section enhancement for the scattering on the largest allowed by kinematics angles and suppression of the forward scattering cross sections in such a way that integrated cross sections remain the same.

\section{Leading orders in the opacity expansion\label{app:derivation}}

In this Appendix, we explicitly derive cross-section modifications by multiple QED re-scattering at the first three orders in the opacity expansion.

The leading in the opacity expansion cross-section correction $\delta \mathrm{d} \sigma^{(1)}$ is expressed in terms of the single-nucleon hard-process cross section $\mathrm{d} \sigma$ as
\begin{equation}
     \delta \mathrm{d} \sigma^{(1)} = \sum \limits_{i} \frac{N_i}{S^i_\perp} \int \frac{\mathrm{d}^2 \vec{q}_{\perp}}{\left(2\pi\right)^2} |v \left( \vec{q}_\perp \right)|^2 \left[ \mathrm{d} \sigma \left( \vec{p}^\prime - \vec{q}_\perp \right)- \mathrm{d} \sigma \left( \vec{p}^\prime  \right) \right], \label{eq:leading_order}
\end{equation}
with the shift of the charged lepton momentum $\vec{p}^\prime$ in the perpendicular direction by $\vec{q}_\perp$ and sum over the layers along the lepton trajectory after the hard interaction, with $N_i$ scattering centers and the cross-sectional area $S^i_\perp$. The correction of Eq.~(\ref{eq:leading_order}) is determined by two contributions: the square of the one-Glauber exchange diagram and interference of the diagram without Glauber interactions with the diagram with two Glauber photons attached to the same spatial point in the nucleus. We present these contributions in the following Fig.~\ref{fig:first_order}.
\begin{figure}[ht]
    \vspace{1.0cm}
    \centering
    \includegraphics[height=0.15\textwidth]{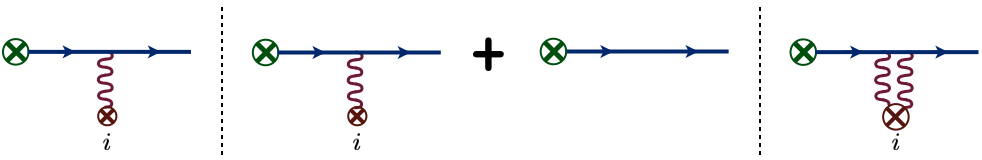}
    \caption{Nonvanishing contributions to the first order in the opacity expansion.}\label{fig:first_order}
\end{figure}

\begin{figure}[h]
    \vspace{1.0cm}
    \centering
    \includegraphics[height=0.475\textwidth]{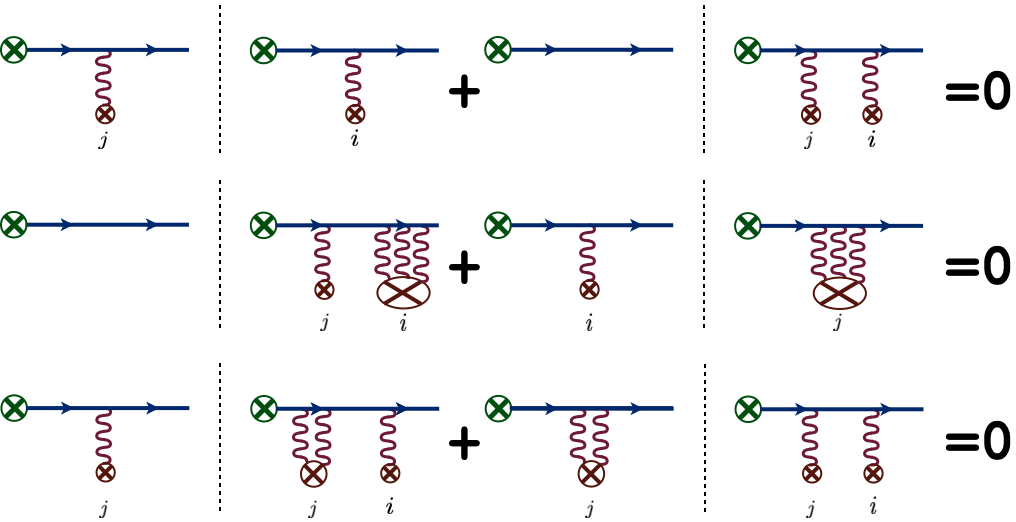}
    \caption{Vanishing in the sum contributions to the second order in the opacity expansion. The same index corresponds to the same interaction point in the coordinate space. The spatial positions of one-, two-, and three-Glauber exchange interactions can be interchanged to account for all topologies.}\label{fig:second_order_zero}
\end{figure}

The second-order contribution in the opacity expansion $\delta \mathrm{d} \sigma^{(2)}$ is expressed as a double sum along the lepton trajectory after the hard interaction
\begin{equation} \label{eq:second_order_+result}
 \delta \mathrm{d} \sigma^{(2)} = \sum \limits_{i>j} \frac{N_i}{S^i_\perp} \frac{N_j}{S^j_\perp} \int \frac{\mathrm{d}^2 \vec{q}^1_\perp}{\left( 2 \pi \right)^2} \frac{\mathrm{d}^2 \vec{q}^2_\perp}{\left( 2 \pi \right)^2} v(\vec{q}^1_\perp)^2 v(\vec{q}^2_\perp)^2 \left[ \mathrm{d} \sigma \left( \vec{p}^\prime - \vec{q}^1_\perp - \vec{q}^2_\perp \right) - 2 \mathrm{d} \sigma \left( \vec{p}^\prime - \vec{q}^1_\perp \right) + \mathrm{d} \sigma \left( \vec{p}^\prime \right) \right].
\end{equation}
We present the non-vanishing contributions to Eq.~(\ref{eq:second_order_+result}) in the following Fig.~\ref{fig:second_order}.
\begin{figure}[ht]
    \vspace{1.0cm}
    \centering
    \includegraphics[height=0.315\textwidth]{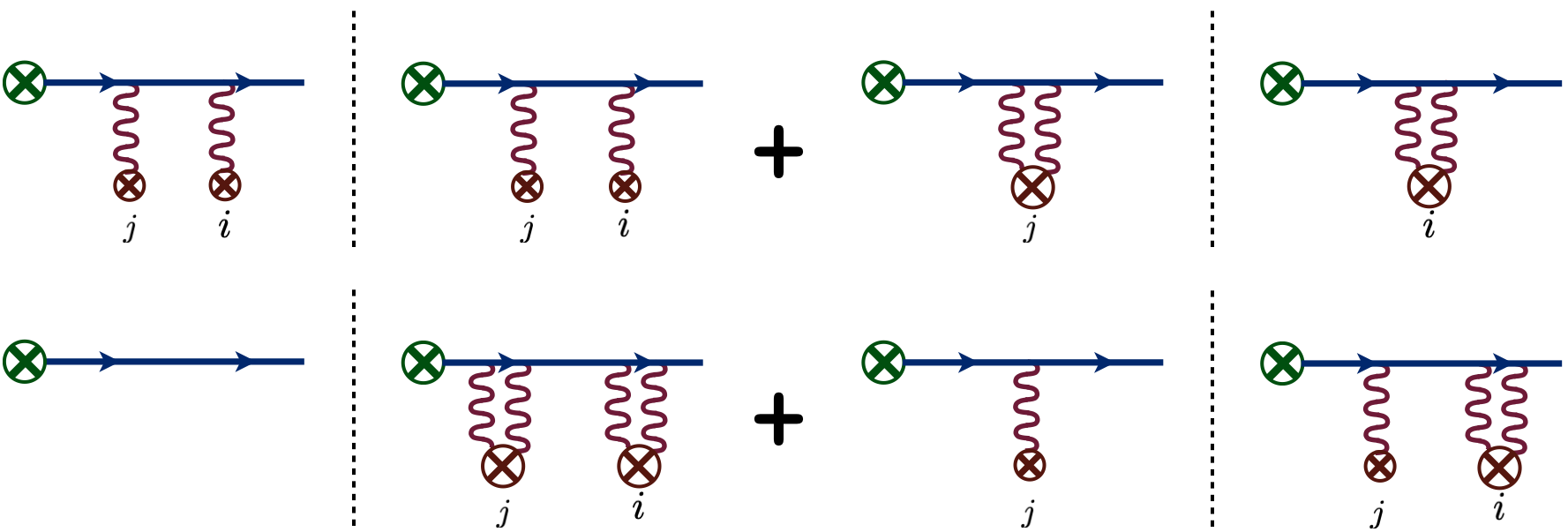}
    \caption{Nonvanishing contributions to the second order in the opacity expansion. The same index corresponds to the same interaction point in the coordinate space. Different indexes represent different points. The spatial positions of one- and two-Glauber exchange interactions can be interchanged to account for all topologies.}\label{fig:second_order}
\end{figure}
At the second order in the opacity expansion, many diagrams, which are proportional to $v \left( 0\right)$ and $v \left( 0\right)^2$, vanish explicitly after the summation over all scattering sites. We show such vanishing contributions in Fig.~\ref{fig:second_order_zero}. In Figs.~\ref{fig:second_order} and~\ref{fig:second_order_zero}, we indicate interaction points as $i$ and $j$ but do not imply the ordering $i>j$ as in Eq.~(\ref{eq:second_order_+result}).

\begin{figure}[ht]
    \vspace{1.0cm}
    \centering
    \includegraphics[height=0.475\textwidth]{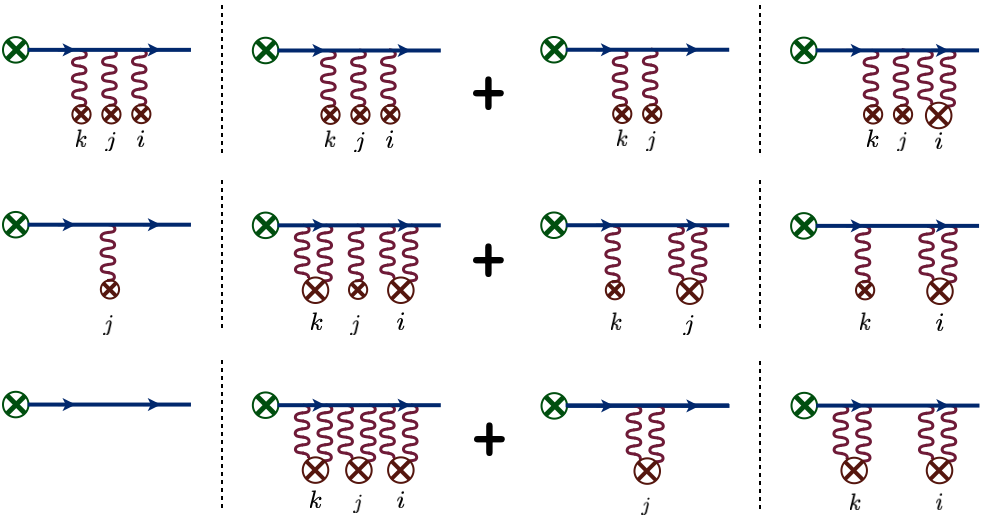}
    \caption{Nonvanishing contributions to the third order in the opacity expansion. The same index corresponds to the same interaction point in the coordinate space. The spatial positions of one- and two-photon exchange interactions can be interchanged to account for all topologies of each diagram.}\label{fig:third_order}
\end{figure}

\begin{figure}[ht]
    \centering
    \includegraphics[height=0.475\textwidth]{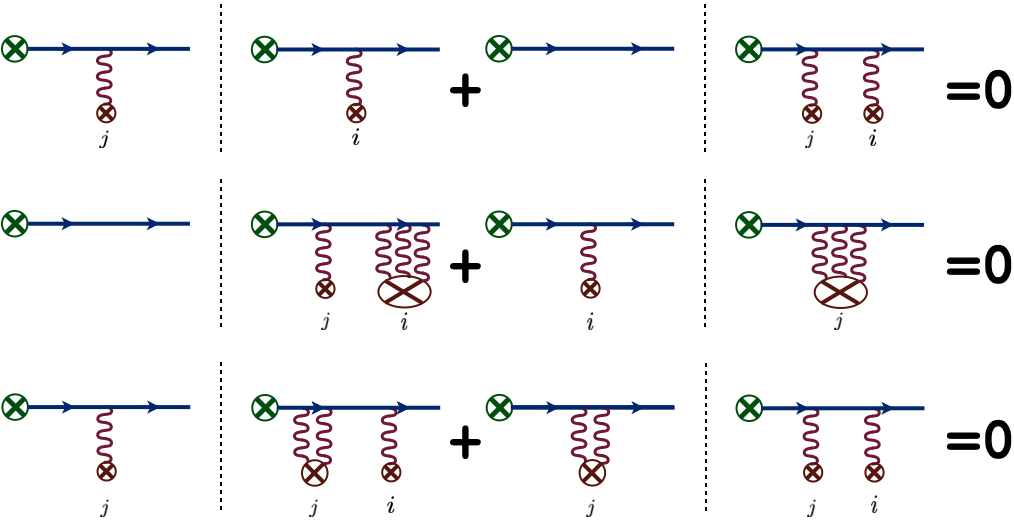}
    \includegraphics[height=0.315\textwidth]{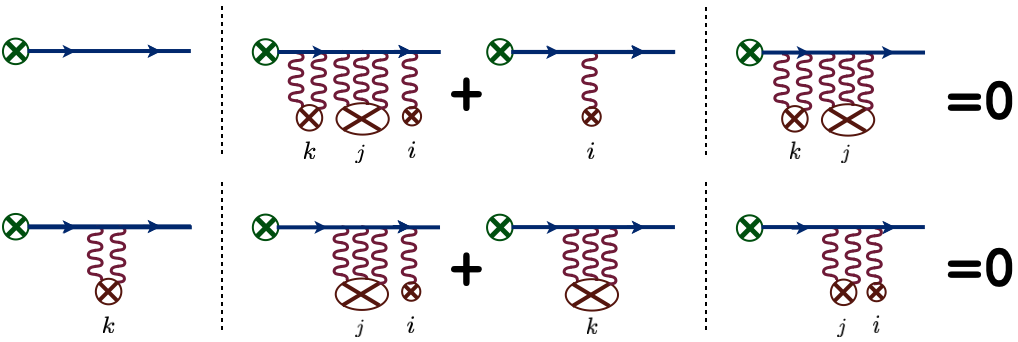}
    \caption{Vanishing in the sum contributions to the third order in the opacity expansion. The same index corresponds to the same interaction point in the coordinate space. The spatial positions of one-, two-, and three-Glauber exchange interactions can be interchanged to account for all topologies.}\label{fig:third_order_zero1}
\end{figure}

\begin{figure}[ht]
    \centering
    \includegraphics[height=0.555\textwidth]{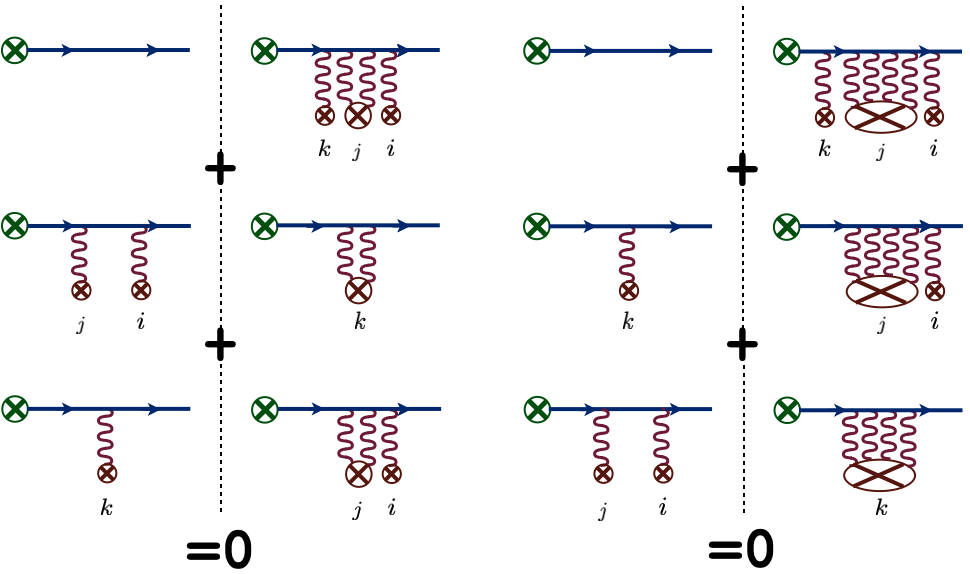} 
    \includegraphics[height=0.555\textwidth]{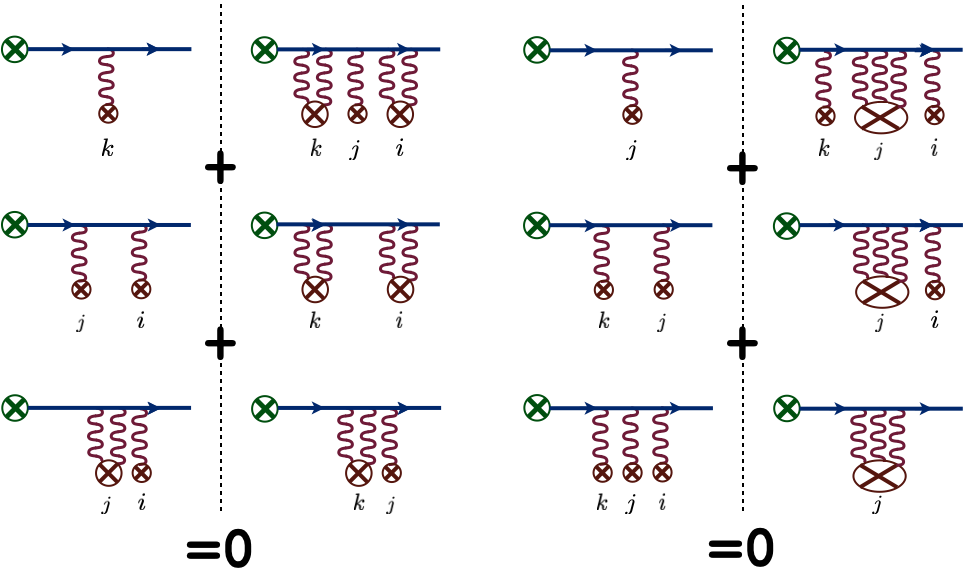}
    \caption{Vanishing in the sum contributions to the third order in the opacity expansion. The same index corresponds to the same interaction point in the coordinate space. The spatial positions of one-, two-, three-, and four-Glauber exchange interactions can be interchanged to account for all topologies.}\label{fig:third_order_zero2}
\end{figure}

The contribution from the third order in the opacity expansion $\delta \mathrm{d} \sigma^{(3)}$ is given by
\begin{align} \label{eq:third_order_+result}
\delta \mathrm{d} \sigma^{(3)} &= \sum \limits_{i > j > k} \frac{N_i}{S^i_\perp} \frac{N_j}{S^j_\perp} \frac{N_k}{S^k_\perp} \int \frac{\mathrm{d}^2 \vec{q}^1_\perp}{\left( 2 \pi \right)^2} \frac{\mathrm{d}^2 \vec{q}^2_\perp}{\left( 2 \pi \right)^2} \frac{\mathrm{d}^2 \vec{q}^3_\perp}{\left( 2 \pi \right)^2} v(\vec{q}^1_\perp)^2 v(\vec{q}^2_\perp)^2 v(\vec{q}^3_\perp)^2 \nonumber \\
&\times \left[\mathrm{d} \sigma \left( \vec{p}^\prime - \vec{q}^1_\perp - \vec{q}^2_\perp - \vec{q}^3_\perp \right) - 3 \mathrm{d} \sigma \left( \vec{p}^\prime - \vec{q}^1_\perp - \vec{q}^2_\perp \right)  + 3 \mathrm{d} \sigma \left( \vec{p}^\prime - \vec{q}^1_\perp \right) - \mathrm{d} \sigma \left( \vec{p}^\prime \right)  \right].
\end{align}
with the non-vanishing diagrams in Fig.~\ref{fig:third_order}.
We also illustrate the vanishing contributions, which are proportional to $v \left( 0 \right)$ and $v \left( 0 \right)^2$, in Figs.~\ref{fig:third_order_zero1} and~\ref{fig:third_order_zero2}.
In Figs.~\ref{fig:third_order},~\ref{fig:third_order_zero1}, and~\ref{fig:third_order_zero2}, we indicate interaction points as $i,~j$ and $k$ but do not imply the ordering $i>j>k$ as in Eq.~(\ref{eq:third_order_+result}).

Noting that only one- and two-Glauber exchange diagrams contribute to Eqs.~(\ref{eq:leading_order})-(\ref{eq:third_order_+result}) with always two Glauber photons in the same site, we generalize the first three orders in the opacity expansion to the well-known all-orders expression in QCD~\cite{Ovanesyan:2011xy} that are used for the derivation of the transverse momentum distributions in Section~\ref{sec:formalism}.

\appendix

\bibliography{paper}{}

\end{document}